\documentstyle[11pt,newpasp,twoside,]{article}
\newcommand{\beq}{\begin{equation}}
\newcommand{\eeq}{\end{equation}}
\newcommand{\bea}{\begin{eqnarray}}
\newcommand{\eea}{\end{eqnarray}}
\newcommand{\ucm}{{\cal D}^{[u]}_t}

\begin{document}

\title{Protostellar Jets Driven by a Disorganized Magnetic Field}
\author{Peter T. Williams}
\affil{Los Alamos National Laboratory, Los Alamos, New Mexico, USA}

\begin{abstract}
We have proposed a viscoelastic model of the Maxwell stresses due to the
disorganized magnetic field in MRI-driven MHD turbulence. Viscoelastic
fluids in the laboratory are known to produce jet-like structures under
the action of a rotating sphere. Here we argue that a similar mechanism
may help explain jets in protostellar systems. Such jets would be driven
not by large-scale organized magnetic fields, but by the mean-field
stresses of small-scale tangled magnetic fields.
\end{abstract}

\section{Introduction}
Broadly speaking, the theory of jet acceleration and collimation has,
over the past decades, progressed from hydrodynamical models to models
incorporating magnetic fields. These models typically involve acceleration
by what may be called ordered, large-scale fields. The long-range 
effects of ordered fields enable, for example, the coupling of a disk
wind or jet to an accretion disk, which is a convenient source of power.

In contrast, the predominant role of magnetic fields in accretion disks
is presumed to be the creation of a large effective turbulent
viscosity, through the action of the Balbus-Hawley magnetorotational
instability (MRI).
The field produced by the MRI, as shown in simulations, has a significant
contribution due to
what might be characterized as a disorganized,
or tangled, field. 
That a tangled field can have such important dynamical consequences as
angular momentum transport in accretion disks leads one to ask what other
consequences such a field might have. 
We propose that one consequence of this tangled field
may be the driving of an axial outflow, {\em i.e.} a jet.

\section{Turbulent Maxwell Stresses}
We break the magnetic field into mean and
fluctuating parts, and the mean is a spatial average over some
sufficiently small length scale 
so as to not blur out dynamically important
details. 
Using an overbar to denote averaging, we write $ B_i = \bar B_i  + B_i'$,
where $B_i$ is a component of the total field, and $B_i'$ is the fluctuating
part. The field $B'$ by definition must have structure on scales equal
to or less than the scale of averaging.
By a ``tangled field'' we mean that the dominant contribution to the
magnetic field is from the small-scale fluctuating component; in other
words, ${\bar B}^2 \ll \overline{B^2}$,
or, equivalently, $\overline{(B')^2} \gg \bar B^2$.
In what follows we will assume that the mean field is negligible, and
thus $B_i = B_i'$. While the mean field may be zero, the mean {\em
stress} will not be for $B' \neq 0$.

The force on a fluid element due to the small-scale component 
of the magnetic field ({\em i.e.}~$B'$) may be written
as the divergence of the turbulent Maxwell stress tensor $\tau_{ij}^{\rm turb}$,
\beq
f_i^{\rm mag} = \partial_j \tau_{ij}^{\rm turb},
\eeq
where the turbulent Maxwell stress tensor is given by
\beq
4 \pi \tau_{ij}^{\rm turb} = \overline{B_i' B_j'} - {1 \over 2} 
\overline{(B')^2} \delta_{ij} = 4 \pi M_{ij} - 2 \pi
M_{kk}\delta_{ij}
\label{eqn:maxwell}
\eeq
and we have defined $4 \pi M_{ij} \equiv \overline{B'_i B'_j}$.
Due to the MRI, the turbulent Maxwell stress tensor in an accretion disk has
a large off-diagonal component, which (along with the Reynolds stress)
is interpreted as an effective viscous
stress; furthermore, there is a large {\em on}-diagonal azimuthal component, which
represents the more-or-less passive advection of the field by the background
shear. The importance of this latter stress is the focus of this paper.

\section{Viscoelastic Models of MHD Turbulence in Disks}
As has been pointed out by Ogilvie (2001) and by Williams (2001),
there is a potentially interesting
analogy between a tangled field in MHD turbulence and tangled polymers in solution.
We emphasize here that in both cases there is a stochastic
element which tends to isotropize the system --- random turbulent motions
of the fluid in the case of MHD turbulence and thermal Brownian
motion in the case of polymers in solution --- and in both cases the action
of large-scale ({\em i.e.} mean-field) shear acts to destroy this isotropy
by aligning the filaments, be they polymers or magnetic field lines.

To be more explicit, we turn to the
vector advection-diffusion equation for the field:
\beq
\partial_t B_i = B_j \partial_j v_i - v_j \partial_j B_i - B_i \partial_j v_j
+ \eta \partial_{jj} B_i.
\label{eqn:vad}
\eeq 
As with the magnetic field, let us decompose the velocity field into
a mean and fluctuating part, $v_i = \bar v_i + v_i'$.
From equation (\ref{eqn:vad}) we may derive the advection equation for the
turbulent Maxwell stress tensor. 
Upon averaging, there will appear  various cross-correlation terms
as well as dissipative terms (abbreviated as $cc.$ and $diss.$), none of which we write explicitly; they will be modeled below.

For simplicity, we write the transport equation for the
tensor $M_{ij}$ instead of for $\tau_{ij}^{\rm turb}$; the latter may easily be obtained
from the former.
We obtain:
\beq
(\ucm M)_{ij}
\equiv
(\partial_t + \bar v_k \partial_k)M_{ij} - (\partial_k \bar v_i)M_{kj} - 
M_{ik}(\partial_k \bar v_j) + 2(\partial_k \bar v_k)M_{ij}= cc. + diss.
\label{eqn:ucm}
\eeq
The notation $\ucm$ is chosen to emphasize that the tensor advective operator
defined above (known as the upper-convected invariant derivative when
the flow is divergenceless) 
is an extension of the more familiar scalar advective derivative,
$\partial_t + v_i \partial _i$, which is often written $D_t$.
If set equal to zero, this operator acting on $M_{ij}$
 would cause the stress in a steady shear flow to
grow without bound. This is the ``elastic'' part of the response of
the magnetized fluid to shear. So long as the dissipation time scale
is shorter than the compressive time scale, the compressibility
effects should be negligible.

Of course the stress does not grow without bound, and this may be
modeled by the inclusion of a relaxation term. The simplest relaxation
term is the stress $M_{ij}$ itself, divided by a relaxation
time $s$. This will appear as a dissipative term, and we need yet
an additional term to act as a source term.

The simplest source term is the effective viscous stress tensor;
this is the so-called Maxwell model (Ogilvie 2001).
Alternatively, one may assume explicitly that random turbulent motions
will tend to isotropize the stress tensor (Williams 2001).
Lastly, one may assume that the Maxwell stresses are produced in proportion
to the Reynolds stresses (this paper). In this last case, we expect that
there may be a useful model for the Maxwell and Reynolds stresses, loosely
derivable from first principles, in which there appear coupled equations
for the evolution of the two stress tensors. We leave this project to
a future paper. The above three models are:
\bea
{\rm Model\ A:}& \qquad s (\ucm M)_{ij} + M_{ij} &= a (\partial_i \bar v_j + \partial_j 
\bar v_i) \\
{\rm Model\ B:}& \qquad s (\ucm M)_{ij} + M_{ij} &= a \delta_{ij} \\
{\rm Model\ C:}& \qquad s (\ucm M)_{ij} + M_{ij} &= a R_{ij}
\eea
Model C is not a complete model in the absence of a further relation to close
the system of equations, but we may still compare its predictions with simulations.
In all three cases, we have two free parameters, namely the relaxation time
$s$ and the coefficient $a$ for the source term that appears on the right.
These determine six parameters, the three diagonal and the three independent
off-diagonal components of the symmetric matrix $M_{ij}$. Note that Ogilvie (2001)
considers a Maxwell model for the full turbulent stress 
(up to magnetic pressure)
$T_{ij} =
M_{ij} - R_{ij}$, where $R_{ij}$ is the Reynolds stress; we will refer to this model
as model $\rm A'$. 

A nice set of simulations of the MRI are those of Hawley, Gammie, \& Balbus~(1995)
and we  use their results
for comparison of the models. It should be noted that 
none of these models as they stand include the effects of a mean field,
whereas all the runs of Hawley et al.~start with a mean seed field to get the
MRI going. Due to flux conservation across the boundaries of the simulation
box, the saturation of the MRI is not wholly independent of these initial
conditions. 

\begin{table}
\caption{Fits of Models A, B, C and ${\rm A}'$ to Hawley et al.}
\begin{tabular}{cccccccccc}
\tableline
qty or index
               & &$M_{ij}$&    A   &    B   &    C   & & &$M_{ij} - R_{ij}$& $A'$   \\
\tableline    	  						    
$rr$           & & ~0.511 & ~0.000 & ~0.249 & ~0.516 & & & -0.452         & ~0.000  \\
$r\theta$      & & -1.000 & -1.000 & -0.741 & -0.921 & & & -1.243         & -1.243  \\
$rz$           & & ~0.003 & ~0.000 & ~0.000 & ~0.007 & & & -0.010         & ~0.000  \\
$\theta\theta$ & & ~3.953 & ~3.961 & ~4.153 & ~4.054 & & & ~3.389         & ~3.386  \\
$\theta z$     & & -0.027 & ~0.000 & ~0.000 & -0.007 & & & -0.040         & ~0.000  \\
$zz$           & & ~0.173 & ~0.000 & ~0.249 & ~0.178 & & & -0.159         & ~0.000  \\
We             & &   --   & ~1.971 & ~2.977 & ~2.036 & & &   --           & ~1.363  \\
b              & &   --   & ~0.030 & ~0.007 & ~0.536 & & &   --           & ~0.037  \\
$\chi^2 / \nu$ & &   --   & ~0.135 & ~0.038 & ~.0003 & & &   --           & ~0.038  \\
\tableline
\tableline
\end{tabular}
\end{table}

In the table below we compare best-fit models with results from
Table 4 of Hawley et al., in which the seed field is azimuthal. 
In our table the parameters We (for Weissenberg) and $b$ are, respectively,
$s$ and $a$ in eqns. (7--9), normalized to the rate of shear. 
Note that models A, B, and C are fit to $M_{ij}$, whereas model ${\rm A}'$ is
fit to $M_{ij} - R_{ij}$. For ease of comparison, we normalize all
data to $-M_{r\theta}$, except for purposes of fitting parameter $b$.
The reduced
chi-square is artificially small in all cases as we have estimated the relative
uncertainty in quantities by using the spatial variance as given in Hawley et al.;
the relative goodness-of-fit is essentially unchanged if one performs a 
simple least-squares fit. 
Similar results
are obtained in comparing the models to Table 2 of that paper, for which the
seed field is vertical; most notable in that case is that the relaxation time
obtained for all models is roughly $2/3$ the relaxation time in the
case of an azimuthal seed field.

As can be seen, the models provide reasonable agreement with simulations,
with model C providing the best fit, although as noted it is not a fully predictive model.
In all three cases the dominant feature is the creation
of a significant streamwise stress $M_{\theta \theta}$ by the advection of 
 $M_{r\theta}$ (or $T_{\theta \theta}$ and $T_{r\theta}$,
respectively, in the case of model~${\rm A}'$, although
it should be noted that the operator $\ucm$ does not
appear in the advection of the Reynolds stress $R_{ij}$).
The creation of a large streamwise stress $M_{\theta \theta}$ is a reliable feature of
any model for MHD turbulence in disks
that includes a treatment of the advection of
turbulent Maxwell stresses, so long as the relaxation time is not very much
faster than the shear time scale. Furthermore, so long as the turbulent
magnetic energy density dominates the turbulent kinetic energy (as it
does here) and the relaxation time is sufficiently long, $M_{\theta \theta}$
will dominate the azimuthal component of the full turbulent stress tensor.

\section{Azimuthal Shear Flow and Hoop-Stresses}
The streamwise stress  $M_{\theta \theta}$ is an azimuthal hoop-stress.
This hoop-stress creates an inwards force
\beq
f_r^{\rm hoop} = - {M_{\theta \theta} \over r},
\eeq
much like a circularly-stretched rubber band.
This is true both in the astrophysical context, as well as in the
 circular shear flow of a viscoelastic fluid in the laboratory.
In fact, a rubber band is an apt
analogy, as a rubber band is itself composed of a linked polymer matrix, and
the stress in anisotropically-stretched 
rubber corresponds to a statistical alignment of these polymers.

One of the more startling laboratory 
demonstrations of the dynamical effects of these hoop
stresses is the reversal of the secondary flow of a viscoelastic fluid in the neighborhood
of a spinning sphere. Through ordinary viscous forces, the sphere induces an azimuthal
shear flow. The inertial term $v_i \partial_i v_j$ in the momentum-conservation equation,
in an ordinary Newtonian fluid,
 flings material out centrifugally 
along the equator; this causes a pressure drop near
the sphere, which sucks fluid in along the poles. If the elasticity of the fluid is
sufficiently strong, however, the above-mentioned hoop-stresses dominate the
centrifugal forces, and the secondary flow is reversed. Fluid is pulled in along
the equator, creating a pressure rise near the sphere; this increased pressure
drives an outward jet-like flow along the axis of rotation of the sphere.

\section{Application to Protostellar Jets}
We propose that essentially the same mechanism responsible for these laboratory
jet-like structures may be responsible for protostellar jet formation, in particular
in those cases in which evidence does not preclude a disk that extends to the 
photosphere of the nascent star. We specifically concentrate on protostellar
jets because we believe the presence of
 a relatively firm central object (as opposed to a black hole) will help
the hoop-stresses drive an outflow, as well as collimate it, which has also
been proposed recently by Li (2002).

For a thick inner disk, we find that if the
disk thickness exceeds the stellar radius by more than a factor of a few, then
the radial force $f_r^{\rm hoop}$ can dominate the
centrifugal, gravitational, and magnetic pressure forces,
depending on the viscosity.
 This force is directed radially inward in
cylindrical, not spherical, coordinates. Following the laboratory analogy,
this force causes a pressure
build-up near the surface of the star that has an outlet in the axial direction
and which, we conjecture, will drive an axial outflow. 
Note that there is net work performed by these hoop-stresses on a fluid element
that spirals in equatorially and is then expelled axially, as there are no
hoop-stresses along the axis.

The current level of understanding of thick disks is far too rudimentary to 
provide anything other than a very rough estimate of the forces and energetics.
In analogy to thin disks, we write the viscosity as
\beq
\nu = \alpha L^2 \left( {\partial \Omega \over \partial \ln r } \right),
\label{eqn:visc}
\eeq
where it is assumed that $\alpha \sim 1$.
Our preliminary results, as given in Williams (2001), are as follows:
We assume that the star is embedded in the thick disk, that is, the thickness
of the disk $H$ is greater than the stellar radius $R_*$. If we assume that
the appropriate $L$ in eqn.~(\ref{eqn:visc}) is $R_*$, then 
it is not clear if the hoop stresses
are sufficient to power an outflow. On the other hand, if we assume
that the appropriate $L$ is the thickness of the disk, then the hoop stresses
are more than sufficient, so long as the jet does not lose too much energy
to viscous dissipation on its way out of the thick disk.



\begin{references}

\reference Hawley, J. F., Gammie, C. F., \& Balbus, S. A. 1995, ApJ, 440, 742

\reference Giesekus, H. 1963 in Fourth International Congress on Rheology, Pt. 1,
ed. E. H. Lee \& A. L. Copley (New York: Wiley), 249

\reference Li, L.-X. 2002, ApJ, 564, 108L, astro-ph/0108469

\reference Ogilvie, G. I. 2001, \mnras, 325, 231, astro-ph/0102245

\reference Thomas, R. H., \& Walters, K. 1964, Quart. Journ. Mech. and Applied Math, 17, 39

\reference Williams, P. T. 2001, {\em xxx.lanl.gov}, astro-ph/0111603

\end{references}
\end{document}